\documentclass[12pt,preprint]{aastex}
\usepackage{natbib}

\newcommand{\agn}{{\small AGN}}

\newcommand{\xmm}{{\it XMM-Newton}}
\newcommand{\rgs}{{\small RGS}}
\newcommand{\chandra}{{\it Chandra}}

\newcommand{\nist}{{\small NIST}}
\newcommand{\nasa}{{\small NASA}}
\newcommand{\fac}{{\small FAC}}
\newcommand{\uv}{{\small UV}}
\newcommand{\osevenplus}{O$^{7+}$}

\newcommand{\nsixplus}{N$^{6+}$}

\newcommand{\cfiveplus}{C$^{5+}$}
\newcommand{\cfourplus}{C$^{4+}$}
\newcommand{\oviii}{\ion{O}{8}}
\newcommand{\ovii}{\ion{O}{7}}
\newcommand{\nvii}{\ion{N}{7}}
\newcommand{\nvi}{\ion{N}{6}}
\newcommand{\cvi}{\ion{C}{6}}
\newcommand{\cv}{\ion{C}{5}}
\newcommand{\cno}{{\small CNO}}
\newcommand{\lya}{Ly$\alpha$}
\newcommand{\lyaone}{Ly$\alpha_1$}
\newcommand{\lyatwo}{Ly$\alpha_2$}
\newcommand{\lyb}{Ly$\beta$}

\newcommand{\lyz}{Ly$\zeta$}

\begin{document}

\title{CNO Line Radiation and the X-ray Bowen Fluorescence Mechanism in
Optically-Thick, Highly-Ionized Media}

\author{Masao Sako\altaffilmark{1}}

\affil{Theoretical Astrophysics,
       California Institute of Technology,
       MC 130-33, Pasadena, CA 91125}

\email{masao@tapir.caltech.edu}

\altaffiltext{1}{\chandra\ Fellow}

\received{February 26, 2003}
\revised{}
\accepted{May 22, 2003}

\slugcomment{Accepted for Publication to ApJ}
\shorttitle{X-ray Bowen Fluorescence}
\shortauthors{Sako}

\begin{abstract}

  Radiative transfer effects due to overlapping X-ray lines in a
  high-temperature, optically-thick, highly-ionized medium are investigated.
  One particular example, where the \ion{O}{8} \lya\ doublet
  ($2~^2P_{1/2;3/2}$ -- $1~^2S_{1/2}$) coincide in frequency with the
  \ion{N}{7} \lyz\ lines ($7~^2P_{1/2;3/2}$ -- $1~^2S_{1/2}$) is studied in
  detail to illustrate the effects on the properties of the emergent line
  spectrum.  We solve the radiative transfer equation to study the energy
  transport of resonance line radiation in a static, infinite, plane-parallel
  geometry, which is used to compute the destruction/escape probabilities for
  each of the lines for various total optical thicknesses of the medium, as
  well as destruction probabilities by sources of underlying photoelectric
  opacity.  It is found that a large fraction of the \oviii\ \lya\ line
  radiation can be destroyed by \nvii, which can result in an reversal of the
  \oviii\ \lya/\nvii\ \lya\ line intensity ratio similar to what may be seen
  under non-solar abundances.  Photoelectric absorption by ionized carbon and
  nitrogen can also subsequently increase the emission line intensities of
  these ions.  We show that line ratios, which are directly proportional to
  the abundance ratios in optically thin plasmas, are not good indicators of
  the true \cno\ abundances.  Conversely, global spectral modeling that
  assumes optically thin conditions may yield incorrect abundance estimates
  when compared to observations, especially if the optical depth is large.
  Other potentially important overlapping lines and continua in the X-ray band
  are also identified and their possible relevance to recent high resolution
  spectroscopic observations with \chandra\ and \xmm\ are briefly discussed.

\end{abstract}


\section{Introduction}

  Radiative transfer effects are capable of producing line spectra that differ
  substantially from those emitted under optically thin conditions.  As
  resonance line photons travel through a medium that is optically thick to
  its own radiation, multiple non-coherent scattering substantially alters the
  line profiles of the emergent spectrum.  Line transfer also affects the
  emergent line intensities through line or continuum absorption, in which
  case the energy is either (1) re-radiated through other discrete transitions
  or (2) is used in heating the ambient gas.  In hydrogen-like ions, for
  example, the upper levels of the higher-series Lyman transitions ($\beta$,
  $\gamma$, $\delta$, etc.) have finite probabilities of decaying to excited
  states.  This implies that the Lyman line photons will be destroyed and
  re-radiated in the Balmer, Paschen, etc. series if the line optical depths
  are sufficiently large.  Ratios of Lyman to Balmer line intensities, for
  example, can then be used to estimate the optical depth, as well as other
  local physical parameters, such as the density and temperature (see, e.g.,
  \citealt{drake80}, and references therein).

  Another well-known example of such an effect is the Bowen fluorescence
  mechanism \citep{bowen34, bowen35}, where it was realized that the
  coincidence in the wavelengths of \ion{He}{2} (He$^{+}$) \lya\ and
  \ion{O}{3} (O$^{2+}$) $2p$ -- $3d$ transitions allows the conversion of
  \ion{He}{2} \lya\ line photons into the upper levels of \ion{O}{3}, which
  then decay to produce lines in the optical/\uv\ band known as the Bowen
  lines at $\lambda \approx 3000 \sim 4000$~\AA.  These line were commonly
  found to be extraordinarily strong in the optical spectra of planetary
  nebulae and Seyfert galaxies.  \citet{unno55} presented the first complete
  analysis of the \ion{He}{2} \lya\ transfer and the conversion efficiencies
  of the Bowen lines, and demonstrated that the observed spectral properties
  can be reasonably well-described by this mechanism.  Subsequently,
  \citet{weymann69} presented a more detailed calculation using an exact form
  of the redistribution function and a realistic ionization balance
  calculation (see, also, \citealt{harrington72, kallman80, eastman85,
  netzer85}).

  In the simplest approximation generally adopted for spectral modeling, one
  assumes that line photons created in the gas escape the medium without any
  further interaction with the constituent atoms in the medium.  This is
  generally referred to as the optically-thin limit or the coronal
  approximation in the context of collisionally ionized plasma.  In this
  limit, the ionization balance is determined solely by the local temperature
  and is assumed to be entirely decoupled from the radiation field.  The
  emission spectra computed under these assumptions seem to apply fairly well
  to the observed X-ray spectra of a wide variety of sources including stellar
  coronae, supernova remnants, hot diffuse gas in starburst galaxies, and the
  intergalactic medium in clusters of galaxies.

  For photoionized plasmas, on the other hand, codes such as Cloudy
  \citep{cloudy} and XSTAR \citep{xstar,bautista01} treat the transfer of
  X-ray lines using either the escape probability method or the assumption of
  complete redistribution.  While these assumptions yield reasonable results
  for {\it isolated} lines with moderate line optical depths ($\tau \la 10$),
  it does not properly describe line scattering when absorption in the damping
  wings becomes important (i.e., for $\tau \ga 100 \sim 1000$ depending on the
  Voigt parameter; see, e.g., \citealt{shine75}).  Such conditions might be
  relevant for a wide variety of astrophysical environment including ionized
  surface layers of accretion disks, extended circumsource regions in \agn,
  and possibly shocked regions in the stellar wind of hot young stars.

  The purpose of this paper is to demonstrate that radiative transfer effects
  can alter intensities of some of the brightest emission lines in the X-ray
  band, in particular, the hydrogen- and helium-like lines from carbon,
  nitrogen, and oxygen.  Non-solar \cno\ abundance ratios have been inferred
  in a number of sources observed with the grating spectrometers on \chandra\
  and \xmm.  The relativistic emission line interpretation of the \xmm\ \rgs\
  spectra of the Seyfert~1 galaxies {\small MCG}$-$6-30-15 and Mrk~766
  presented by \citet{branduardi01}, for example, requires the nitrogen
  emission lines, which are presumably formed in an optically thick accretion
  disk, to be much stronger than that of the oxygen line assuming solar
  abundance ratios (see also, \citealt{mason03}).  Similarly,
  \citet{kinkhabwala02} conclude that nitrogen in the extended circumnuclear
  regions in the Seyfert~2 galaxy {\small NGC}~1068 is overabundant by a
  factor of $\sim 3$ based on measurements of the hydrogen- and helium-like
  line intensities with the \rgs\ (see also, \citealt{brinkman02, ogle03}).
  \citet{jimenez02} also report abnormally high nitrogen line intensities
  relative to those of oxygen in the \rgs\ spectrum of the low-mass X-ray
  binary Her~X-1, and interpreted as an overabundance of nitrogen due to
  H-burning by the \cno-cycle.  All of the modeling, however, are based on
  rather simple models that do not include detailed line transfer.  We show
  that the effects of line overlap alone can explain some of the observed
  anomalies, in particular the strength of the nitrogen lines, if the optical
  depth is large, and that detailed transfer models are required to infer
  accurate values for the \cno\ abundances with X-ray spectroscopic data.  We
  note, however, that the theoretical model presented here is highly idealized
  in order to simply illustrate the general effects and describe the relevant
  spectroscopic details, and is certainly not adequate for quantitative
  comparisons with observations of complicated systems.

  The paper is organized as follows.  In \S2, we describe the spectroscopic
  properties of the frequency range near the \oviii\ (\osevenplus) \lya\
  doublet transitions, which contains a pair of overlapping \nvii\ (\nsixplus)
  \lyz\ lines and underlying sources of continuum opacity, which is usually
  dominated by K-shell absorption of H- and He-like carbon (\cfourplus\ and
  \cfiveplus) and He-like nitrogen (\nsixplus).  We compute the destruction
  probabilities of \oviii\ \lya\ due to line absorption by the \nvii\ \lyz\
  transition and continuum absorption by carbon and nitrogen for temperatures
  ranging from cool photoionized media ($kT \sim 10~\rm{eV}$) to hot
  collisionally ionized plasmas ($kT \sim 1~\rm{keV}$), which is described in
  \S3.  We demonstrate that the wavelength coincidence allows efficient
  conversion of \oviii\ \lya\ line radiation into the \nvii\ lines, and may be
  misidentified as a \cno\ abundance anomaly (\S4).  Finally, in \S5, a few
  other potentially important overlaps are identified and their effects on the
  global X-ray spectrum under optically thick conditions are briefly
  discussed.

\section{Spectroscopy}

\begin{figure}
  \includegraphics[scale=0.9]{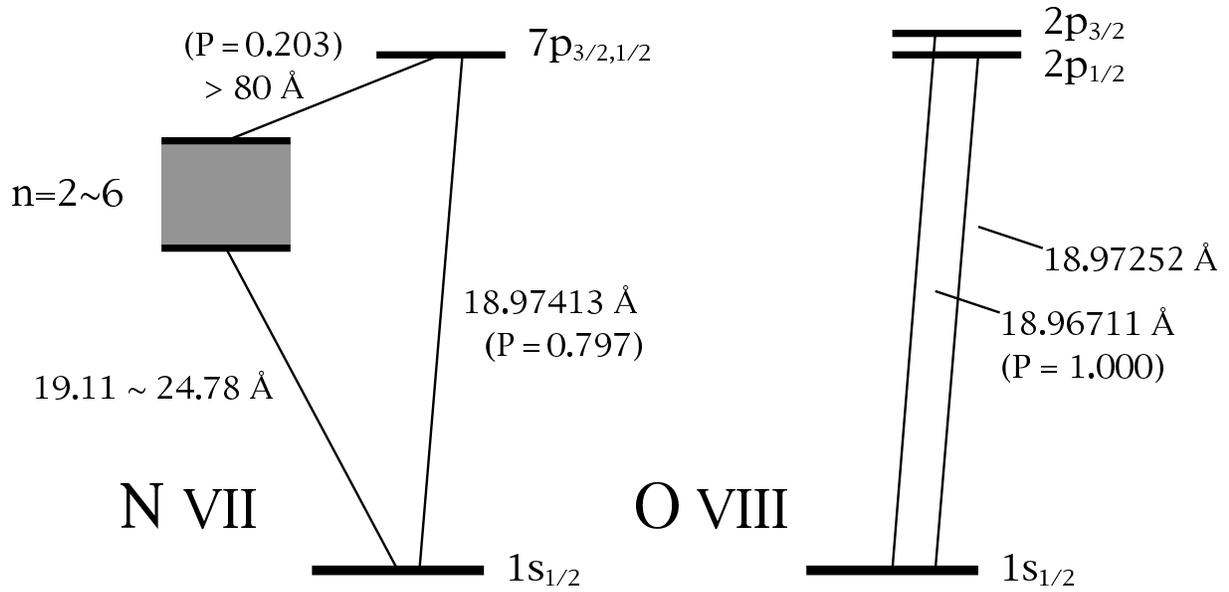}
  \caption{Partial Grotrian diagrams of the \oviii\ and \nvii\ transitions
  relevant for the X-ray Bowen fluorescence mechanism.  The probabilites for
  each channel $P \equiv 1 - \epsilon$ are indicated.}  \label{fig:grot}
\end{figure}

  In the non-relativistic, hydrogenic approximation, the transition energy of
  the \oviii\ \lya\ line ($n = 2 \rightarrow 1$) coincides exactly with that
  of the \nvii\ \lyz\ ($n = 7 \rightarrow 1$) line; i.e., $Z_a^2(1-1/n^2) =
  Z_b^2(1-1/m^2)$ holds exactly for $Z_a = 8$, $n = 2$ and $Z_b = 7$, $m = 7$.
  More precisely, the \oviii\ \lya\ transition is a doublet at wavelengths
  $\lambda_{\rm{Ly}\alpha_1} = 18.96711$~\AA\ ($653.680~\rm{eV}$) and
  $\lambda_{\rm{Ly}\alpha_2} = 18.97251$~\AA\ ($653.494~\rm{eV}$) that
  correspond to decays to the ground state from the $^2P_{3/2}$ and
  $^2P_{1/2}$ levels, respectively.  The oscillator strength of the \lyaone\
  line is twice that of the \lyatwo\ line.  The \ion{N}{7} \lyz\ transition is
  also a doublet with $\lambda_{\rm{Ly}\zeta_1} = 18.97411$~\AA\
  ($653.439~\rm{eV}$) and $\lambda_{\rm{Ly}\zeta_2} = 18.97418$~\AA\
  ($653.436~\rm{eV}$).  Since the wavelength difference of $0.07$~m\AA\ is
  much too small to be resolved at any reasonable temperature where \nsixplus\
  can exist, we simply treat them as a single line with an
  oscillator-strength-averaged wavelength of $\lambda = 18.97413$~\AA\
  ($653.438~\rm{eV}$; see, Figure~\ref{fig:grot}).  The \oviii\ \lya\ and
  \nvii\ \lyz\ transition wavelengths are adopted from \citet{johnson85} and
  \citet{garcia65}, respectively, which include reduced-mass and quantum
  electrodynamical corrections to the eigenvalues of the Dirac
  equation\footnote{see, also the \nist\ Atomic Spectra Database --
  http://physics.nist.gov/cgi-bin/AtData/main\_asd}.  Oscillator strengths and
  radiative decay rates are calculated using the Flexible Atomic Code
  (\fac\footnote{ftp://space.mit.edu/pub/mfgu/}; \citealt{fac}).  All of the
  relevant atomic parameters are listed in Table~\ref{tbl1}.

\begin{figure}
  \includegraphics[scale=0.7,angle=-90]{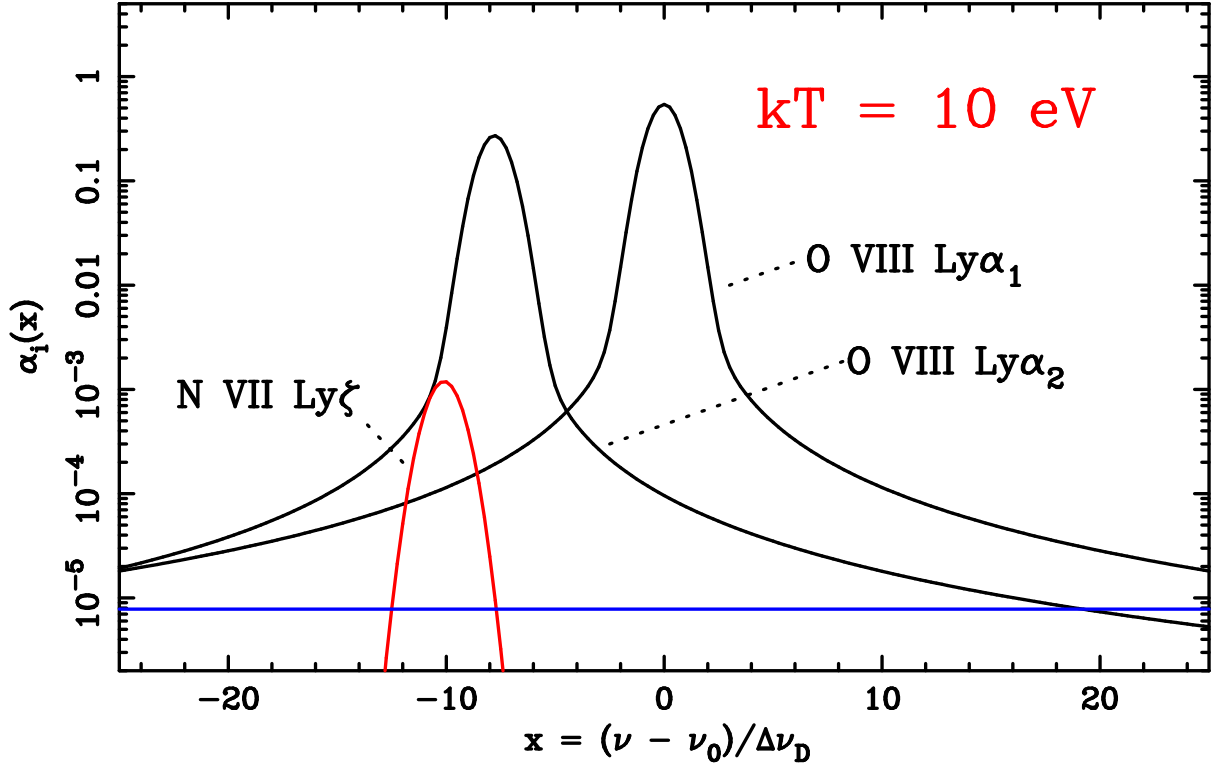}
  \caption{Normalized cross sections of as functions of the frequency for
  \oviii\ \lya\ and \nvii\ \lyz\ at a temperature of $10~\rm{eV}$.  The
  frequency is expressed in multiples of the Doppler width of the \oviii\
  \lyaone\ line.  The nitrogen to oxygen abundance ratio is assumed to be
  0.13.  The continuum absorption cross section is set to the value of $\beta$
  that corresponds to a column density of $\tau_c = 0.1~\tau_{\rm{O~VIII}}$.}
  \label{fig:cross10}
\end{figure}

\begin{figure}
  \includegraphics[scale=0.7,angle=-90]{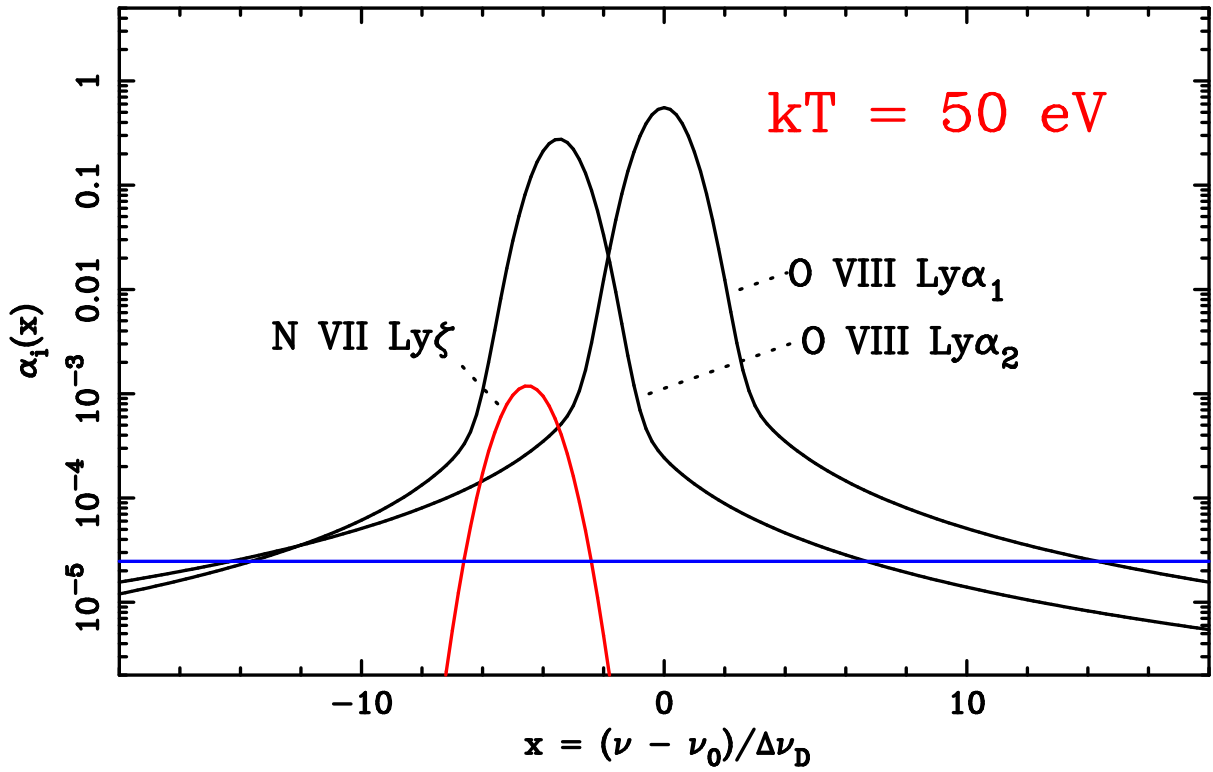}
  \caption{Same as in Figure~\ref{fig:cross10} at a temperature of $kT =
  50~\rm{eV}.$} \label{fig:cross50}
\end{figure}

  The relative cross sections assuming solar abundances
  ($A_{\rm{N}}/A_{\rm{O}}$ = 0.13; \citealt{anders89}) and a temperature of
  $kT = 10 ~\rm{eV}$, typical for a photoionization-dominated medium at this
  level of ionization, are shown in Figures~\ref{fig:cross10}.  At this
  temperature, the \nvii\ \lyz\ lies approximately 2~Doppler widths away from
  the \oviii\ \lyatwo\ line and 10 Doppler widths from the \lyaone\ line.  The
  cross section is almost three orders of magnitude lower than that of the
  \oviii\ lines, which implies that significant scattering between the lines
  occur when the total optical depth is $\ga 10^3$.  Similarly, at a
  temperature of $kT = 50~\rm{eV}$, the \nvii\ \lyz\ lies $\sim 1$~Doppler
  width away from the \oviii\ \lyatwo\ line and 4 Doppler widths from the
  \lyaone\ line, as shown in Figure~\ref{fig:cross50}.

  The emission coefficient in this spectral range is dominated almost entirely
  by the \oviii\ \lya\ lines.  The intrinsic \nvii\ \lyz\ emission coefficient
  is approximately two to three orders of magnitude lower in both
  collision-dominated and recombination-dominated plasmas with solar
  $A_{\rm{N}}/A_{\rm{O}}$ abundance ratio.  If the medium is photoionized by
  some external source of continuum radiation, photoexcitation may enhance the
  line emissivities of the \nvii\ lines, but this is important only in the
  surface layer within $\tau \la 10$, as external radiation will not be able
  to penetrate deeper into the medium.  Since we are primarily interested in
  the properties of the Bowen mechanism at large optical depths, we ignore the
  \nvii\ \lyz\ intrinsic source term entirely.

  Absorption of an \oviii\ line photon by \nvii\ is followed by one of the
  following two processes: (1) re-emission in the \ion{N}{7} \lyz\ line, which
  results in a pure scattering event or (2) destruction of the line through
  cascades via the upper levels ($n = 2 - 6$), followed eventually by decay to
  the ground level.  Collisional excitation and de-excitation from these
  levels are important at only extremely high electron densities ($n_{\rm
  e,crit} \sim 10^{21}~\rm{cm}^{-3}$), and so we do not consider these
  processes here.  We note, however, that thermalization of the line photons
  via collisions may occur at sufficiently high density and high optical
  depth, such that $n_e \langle{N}\rangle_{\rm line} \ga n_{\rm e,crit}$,
  where $\langle{N}\rangle_{\rm line}$ is the mean number of scatterings
  experienced by the line photons as they propagate through the medium.

  The branching ratios for the two possible processes can be determined from
  the radiative decay rates from the $7p$ level to all possible levels.  We
  use radiative rates calculated with \fac\ except for the two-photon decay
  rates, which are adopted from \citet{drake86}.  The branching ratios for
  processes (1) and (2) are found to be 0.797 and 0.203, respectively.  The
  final fate of the 20.3\% of the line photons that initially decay to an
  upper level can be understood by computing the cascade matrices; quantities
  that represent the probabilities of one state decaying to another state via
  all possible channels \citep{seaton59}.  We find, for example, that 58\% of
  them produce the Balmer, Paschen, etc.\ lines and eventually result in
  two-photon emission from the $2~^2S_{1/2}$ level, while the remaining 42\%
  produce the Lyman lines in the soft X-ray band.

  Continuum opacity from ions that coexist with \oviii\ is dominated primarily
  by \cv\ (ionization potential $\chi = 31.622$~\AA), \cvi\ ($\chi =
  25.303$~\AA), and \nvi\ ($\chi = 22.458$~\AA) for both collisionally and
  photoionized plasmas in ionization and thermal equilibrium.  The threshold
  wavelength of \nvii\ ($\chi = 18.587$~\AA) lies shortward of the \oviii\
  \lya\ line.  Lower charge states of carbon and nitrogen may contibute very
  small amounts of continuum opacity as well.  For solar abundances, the
  typical value for the combined continuum opacity relative to the \oviii\
  \lya\ line opacity is $\sim 10^{-4}$, but can range anywhere from a
  $\rm{few} \times 10^{-6}$ to as high as $\sim 10^{-3}$ depending on the
  level of ionization and the temperature.  In ionization equilibrium, a
  carbon or nitrogen ion that absorbs the \oviii\ \lya\ line will be
  photoionized, and will eventually recombine to form emission lines and
  continua.  In a recombination-dominated plasma at a temperature of $kT \sim
  100~\rm{eV}$, for example, calculations show that approximately 50\% of the
  total number of recombinations produce one of the Lyman lines.

  Under most circumstances, the Compton cross section is roughly two orders of
  magnitude below the continuum absorption cross section.  This implies that
  lines photons are preferentially absorbed before they are Compton scattered
  and, therefore, continuum absorption is the dominant mechanism that competes
  with the line processes.  Compton scattering effects will, however, be
  important in extremely highly-ionized regions, where only trace abundances
  of bound ions can survive or in low-metalicity plasmas with abundances $\la
  1$\% solar.  These will be ignored entirely in the present work.

\section{Computational Method}

  To illustrate the importance of line overlap, we assume an infinite,
  symmetric slab with a pre-specified source distribution of line photons.  In
  reality, dynamical effects, such as velocity gradients and turbulence, as
  well as inhomogenieties that exist, for example, in accretion disks
  complicate the problem, but we ignore them in the present study.  Although
  resonance line scattering, in general, produces polarized light, we assume
  that the radiation field is unpolarized (cf., \citealt{lee94}).  To simpify
  the computational procedure even further, the medium is assumed to be
  isothermal with a constant level of ionization.  We solve the radiative
  transfer equation in the frequency range near the \oviii\ \lya\ lines by
  properly accounting for scattering and absorption by the various overlapping
  lines, as well as the presence of a background continuum opacity source.
  Using the emergent intensity, we then compute the escape probabilities of
  the original \oviii\ \lya\ line photons and conversion efficiencies into the
  \nvii\ lines and photoelectric absorption as functions of the total optical
  depth of the slab.

  The radiative transfer equations for rays propagating in the $+\mu$ and
  $-\mu$ directions ($\mu~\equiv~cos~\theta$, where $\theta$ is the angle from
  the normal $z$-axis) in a plane-parallel medium can be written as,
\begin{equation}
  \label{eq:rte1}
  \pm \mu \frac{\partial I(z,\nu,\pm\mu)}{\partial z} =
    \sum_i k_i(\nu)[S_i(z,\nu) - I(z,\nu,\pm\mu)]
\end{equation}
  where $\theta$ is the angle measured from the $z$-direction, which we define
  to be normal to the slab.  The right-hand-side is summed is over all
  possible sources of opacity and emissivity in the slab.  We express the
  frequency in terms of the frequency shift from the reference line center at
  frequency $\nu_0$ in units of the Doppler width $\Delta \nu_D$, i.e., $x =
  (\nu - \nu_0)/\Delta \nu_D$, and write the line absorption coefficient
  (cm$^{-1}$) as $k_i(x) = k_{iL} \phi_a(x)$.

  Since most X-ray resonance lines have large radiative decay rates ($A_r \ga
  10^{12} ~\rm{s}^{-1}$), the population of the upper level is usually
  negligible compared to that of the ground state.  Denoting the transition
  oscillator strength as $f_{ul}$ and the lower-level density as $n_i$
  ($\rm{cm}^{-3}$), we can then write,
\begin{equation}
  \label{eq:kL}
  k_{iL} = \frac{\pi e^2}{m_e c} \frac{f_{ul} n_i}{\Delta \nu_D},
\end{equation}
  and $\phi_a(x)$ is the normalized Voigt profile,
\begin{equation}
  \label{eq:voigt}
  \phi_a(x) = \frac{H(a,x)}{\sqrt{\pi}} =
    \frac{a}{\pi^{3/2}} \int^{\infty}_{-\infty}
    \frac{e^{-t^2} dt}{(x-t)^2 + a^2}.
\end{equation}
  Adopting the notation of \citet{hummer80}, we define $\tau$ to be the mean
  optical scale, which increases in the negative $z$-direction through the
  slab.  Therefore, $d\tau = - k_L dz$, which is related to the line center
  optical depth $\tau_0$ by, $\tau_0 = \phi_a(0) \tau$.

  We assume that the continuum is purely absorbing with zero local emissivity.
  The source function then contains only the line terms, and can be written as
  \citep{hummer69},
\begin{equation}
  S_{iL}(\tau,x) = \frac{1 - \epsilon_i}{\phi_{i,a}(x)} \int_{-\infty}^{+\infty}
    R(x,x') J(\tau,x') dx' + G_i(\tau),
\end{equation}
  where the first term represents the scattering term and $G_i(\tau)$ is the
  intrinsic source term.  The quantity $\epsilon_i$ is the destruction
  probability of the line per scattering (i.e., $\epsilon = 0.203$ for the
  \nvii\ \lyz\ line) and $J(\tau,x)$ is the angle-averaged intensity,
\begin{equation}
  J(\tau,x) = \frac{1}{2} \int_{-1}^{+1} I(\tau,x,\mu)d\mu.
\end{equation}
  The scattering term is computed assuming partial redistribution in a Voigt
  profile, in which photons scatter coherently in the atoms' frame.  Compared
  to the assumption of complete redistribution, this provides a more accurate
  representation of line scattering in an optically thick medium where
  scattering in the damping wings is important.  The frequency redistribution
  function $R(x,x')$ is defined as the probability that a photon of frequency
  $x$ absorbed by an ion is re-emitted at frequency $x'$, and can be written
  as,
\begin{equation}
  R(x,x') = \frac{1}{\pi^{3/2}} \int^{+\infty}_{-\infty} e^{-u^2}
    \left\{\tan^{-1}\left[\frac{{\rm{max}}(x,x') + x}{a}\right] -
          \tan^{-1}\left[\frac{{\rm{min}}(x,x') - x}{a}\right] \right\} du,
\end{equation}
  which was first derived by \citet{henyey41} (see, also \citealt{hummer62},
  and references therein).  Accounting for the angular redistribution makes
  negligible difference \citep{milkey75} and is, therefore, ignored.  The
  intrinsic source term is frequency independent and can be produced, for
  example, through collisional excitation and or radiative cascades following
  recombination.  The exact nature of the excitation mechanism is not
  important for our present purposes.

  The transfer equation for multiple overlapping lines can then be written as,
\begin{equation}
  \label{eq:rte2}
  \pm \mu \frac{dI_\nu(x,\mu,\tau)}{d\tau} =
    \sum_{i} \left[ I(x,\mu,\tau) \alpha_{i}(x) +
    (1 - \epsilon_{i}) \frac{\alpha_{i}(x)}{\phi_a(x)}
    \int^{\infty}_{-\infty}
    R_{i}(x', x) J(\tau,x') dx' + \alpha_{i}(x) G_{i}(\tau) \right],
\end{equation}
  where $\alpha_i(x)$ represent the absorption coefficients normalized to that
  of the reference line at $x = 0$, and are assumed to be independent of
  depth.  Therefore, the first, second, and third terms on the right hand side
  of Eq.~\ref{eq:rte2} represent absorption, scattering, and intrinsic
  emission, respectively.

  The transfer equation is solved using the Feautrier method
  \citep{feautrier64}, which requires a discretized version of
  Eq.~\ref{eq:rte2} in angular, frequency, and optical depth variables.
  Defining the total opacity $\phi_{\rm{tot}}(x) = \sum\phi_{i,a}(x) + \beta$,
  where $\beta$ is the continuum opacity relative to the reference line
  opacity ($\beta = k_C/k_{L}$) assumed to be independent of frequency across
  the line, we define a mean-intensity-like quantity,
\begin{equation}
  j_{\mu\nu} = \frac{1}{2}[I(x,+\mu,\nu) + I(x,-\mu,\nu)],
\end{equation}
  and rewrite Eq.~\ref{eq:rte2} as a second order differential equation,
\begin{equation}
  \label{eq:rte3}
  \mu^2 \frac{d^2j_{\mu\nu}}{d\tau^2} = \phi^2_{\rm{tot}}(x) j_{\mu\nu} +
    \phi_{\rm{tot}}(x) \sum_{i=0} \alpha_i(x) S_i(\tau, x).
\end{equation}
  The presence of temperature and ion abundance gradients will introduce an
  additional term on the right hand side, which is ignored for simplicity.  We
  assume a symmetric slab and specify boundary conditions on the surface and
  at the midplane of the slab.  Assuming no incident radiation on either
  boundaries, they can be written as,
\begin{equation}
  \label{eq:bc_top}
  \mu \left. \frac{dj_{\mu\nu}}{d\tau} \right \vert_{\tau = 0} = j_{\mu\nu}
\end{equation}
  on the surface and,
\begin{equation}
  \label{eq:bc_mid}
  \mu \left. \frac{dj_{\mu\nu}}{d\tau} \right \vert_{\tau = \tau_{\rm{tot}}/2} = 0
\end{equation}
  at the midplane.  These boundary conditions are also discretized using a
  second order method described by \citet{auer67}.  The frequency integral is
  represented as a quadrature sum using Simpson's formula with $N_x = 100 -
  200$ points depending on the temperature and the total optical depth of the
  slab.  We use a 2-point Gaussian quadrature for the angular integral in the
  interval $0 \leq \mu \leq 1$ ($N_\mu = 2$).  The optical depth is
  discretized into 12 points per dex.  Extensive testing shows that the above
  discretization scheme yields reliable results.  Energy conservation (see,
  e.g., \citealt{hummer80}) is satisfied to within $\sim 2$\% for most cases
  studied in this paper, and is no larger than $\sim 5$\% even in the worst
  cases.

\section{Conversion Efficiencies}

  We solve the transfer equation for a source function distributed uniformly
  within a slab of ionized material for a range of total slab thicknesses.  We
  define the \oviii\ \lyaone\ line to be our reference line and measure
  frequency shifts in units of its Doppler width.  The emissitivities of the
  \oviii\ \lyaone\ and \lyatwo\ lines are assumed to be proportional to their
  respective statistical weights of the upper levels, i.e.,
  $G_{\rm{Ly}\alpha_1}(\tau) = 2 G_{\rm{Ly}\alpha_2}(\tau)$, which is
  approximately correct for both collision-dominated and
  recombination-dominated line emission.  The absolute numerical values for
  the source functions are not important for the present purposes, and without
  any loss of generality, we can write,
\begin{equation}
  \label{eq:src_norm}
  \sum_i \int_0^{\tau_{\rm{tot}}} d\tau_i G_i(\tau) \equiv 1.
\end{equation}
  The solution to Eq.~\ref{eq:rte3}~--~\ref{eq:bc_mid} yields the
  frequency-dependent intensity at each optical depth and angular points,
  which can then be used to compute the mean number of scatterings for each of
  the lines, their mean optical scales, and the escape probabilities.

  Given the normalization condition (Eq.~\ref{eq:src_norm}), the fraction of
  the original photons that are absorbed by a transition with destruction
  probability $\epsilon_i > 0$ can be expressed in terms of the mean number of
  scatterings \citep{hummer69, hummer80}.  The fraction absorbed and destroyed
  by line $i$ can be written as,
\begin{equation}
  f_{iL} = \frac{\epsilon_i}{1 - \epsilon_i}\langle N \rangle_i =
           \frac{\epsilon_i}{1 - \epsilon_i} \int_0^{\tau_{\rm{tot}}} d\tau_i
           \int_{-\infty}^{\infty} \phi_a(x) S_{iL}(\tau,x) dx,
\end{equation}
  where $\langle N \rangle_i$ is the mean number of scatterings.  Similarly,
  the fraction of photons absorbed in the continuum can be expressed in terms
  of the mean optical scale $\langle l \rangle$ \citep{ivanov73}, i.e.,
\begin{equation}
  f_C = \beta \langle l \rangle = \beta \int_0^{\tau_{\rm{tot}}} d\tau
        \int_{-\infty}^{\infty} J(\tau,x) dx.
\end{equation}
  Finally, the fraction of photons that escape through both faces of the slab
  is simply,
\begin{equation}
  f_{\rm{esc}} = \int_{-\infty}^{\infty} dx \int_{0}^{1} I(\tau = 0,
                 x,\mu)~\mu~d\mu,
\end{equation}
  where $I(\tau = 0,x,\mu)$ is the emergent intensity on the surface.

\begin{figure}
  \includegraphics[scale=0.7,angle=-90]{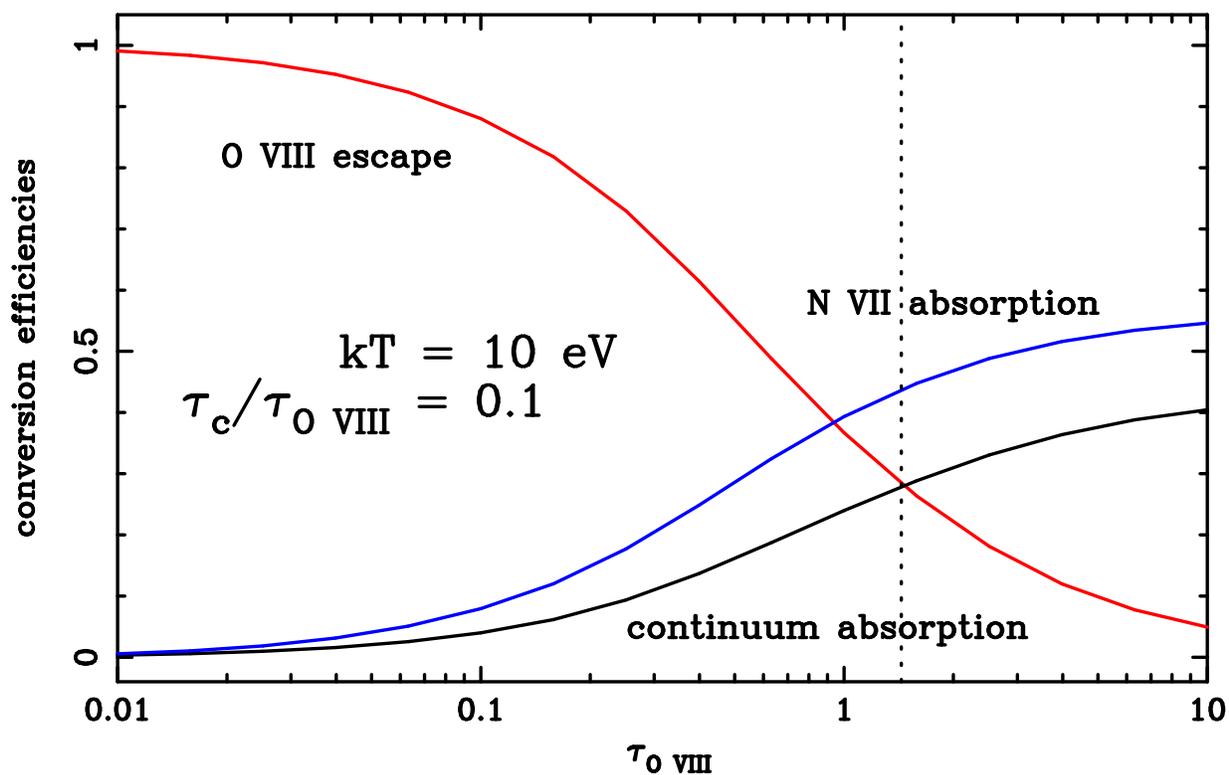}
  \caption{Fraction conversion efficiencies at $kT = 10~\rm{eV}$ as a function
  of the total optical depth of the slab expressed in units of the threshold
  photoelectric opacity of \oviii.  The underlying continuum opacity is fixed
  to be $\tau_c = 0.1~\tau_{\rm{O~VIII}}$.  The freqency dependences of the
  line cross sections are shown in Figure~\ref{fig:cross10}.  The vertical
  dotted line denotes where the \oviii\ \lyaone\ line-center optical depth is
  $10^4$.}
  \label{fig:yield10}
\end{figure}

\begin{figure}
  \includegraphics[scale=0.7,angle=-90]{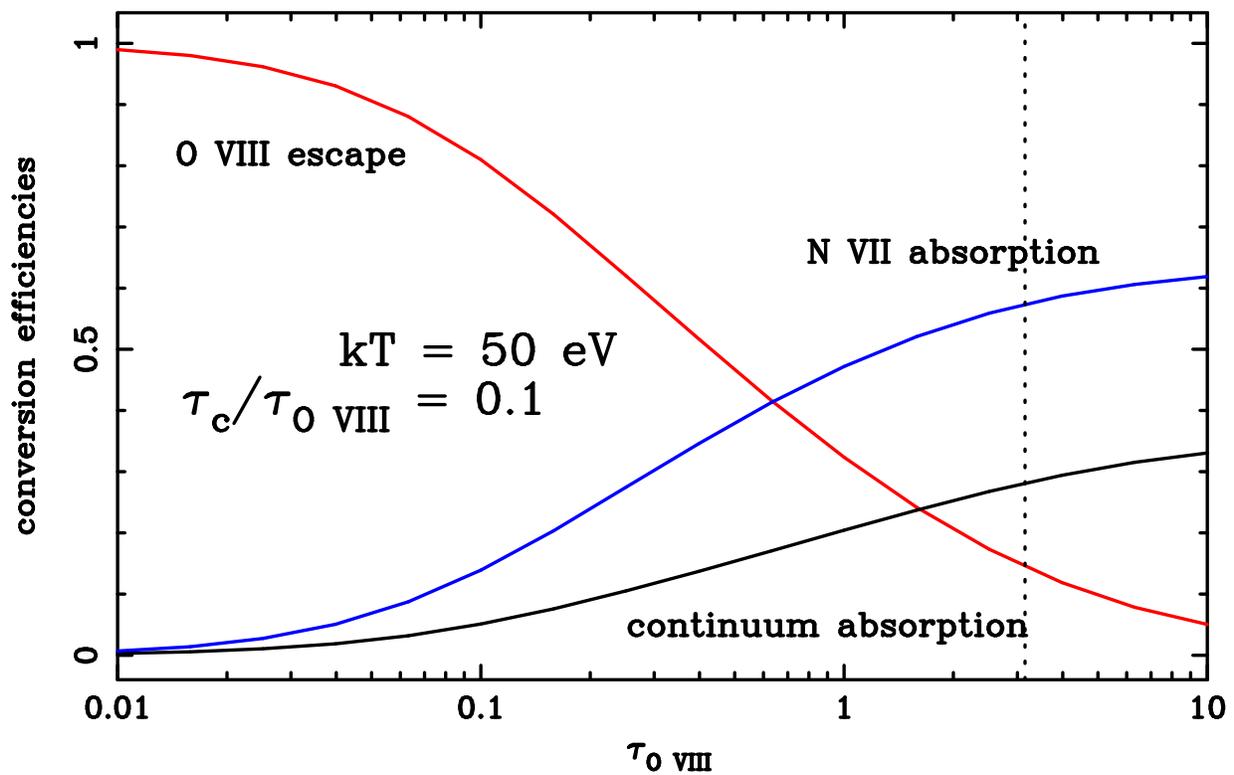}
  \caption{Same as in Figure~\ref{fig:yield10} at $kT = 50~\rm{eV}$.  The
  freqency dependences of the line cross sections are shown in
  Figure~\ref{fig:cross50}.}  \label{fig:yield50}
\end{figure}

  Fractional conversion efficiencies as a function of the total threshold
  optical depth at the \oviii\ edge ($\tau_{\rm{O~VIII}}$) at temperatures of
  $kT = 10 ~\rm{eV}$ and $50 ~\rm{eV}$ are shown in Figures~\ref{fig:yield10}
  and \ref{fig:yield50}, respectively.  The underlying continuum opacity is
  assumed to be $0.1~\tau_{\rm{O~VIII}}$.  At $kT = 10 ~\rm{eV}$, the
  line-center cross section of the \oviii\ \lyaone\ line is $6.9 \times
  10^{-16}~\rm{cm}^2$ and is a factor of $7.0 \times 10^3$ times larger than
  the corresponding continuum absorption cross section at threshold, which is
  assumed to be $9.9 \times 10^{-20}~\rm{cm}^2$ for \oviii\ \citep{verner96}.
  For $\tau_{\rm{O~VIII}} = 1$ and $kT = 10 ~\rm{eV}$
  (Figure~\ref{fig:yield10}), only $\sim 40$\% of the \oviii\ \lya\ photons
  escape the medium, while $\sim 40$\% are absorbed by \nvii\ and $\sim 20$\%
  are absorbed by the background continuum.  The total line-center opacity in
  the \oviii\ \lyaone\ line is $6.9 \times 10^3$.  We find the mean number of
  scatterings for the \oviii\ \lyaone\ line to be $2.3 \times 10^3$, which
  implies that thermalization through particle collisions is relevant if the
  gas density is higher than $\sim 10^{18}~\rm{cm}^{-3}$ (see \S2).  At a
  temperature of $kT = 50 ~\rm{eV}$, the overlap between the \oviii\ and
  \nvii\ lines is larger, and the conversion to \nvii\ is enhanced relative to
  the amount absorbed in the continuum.

\begin{figure}
  \includegraphics[scale=0.7,angle=-90]{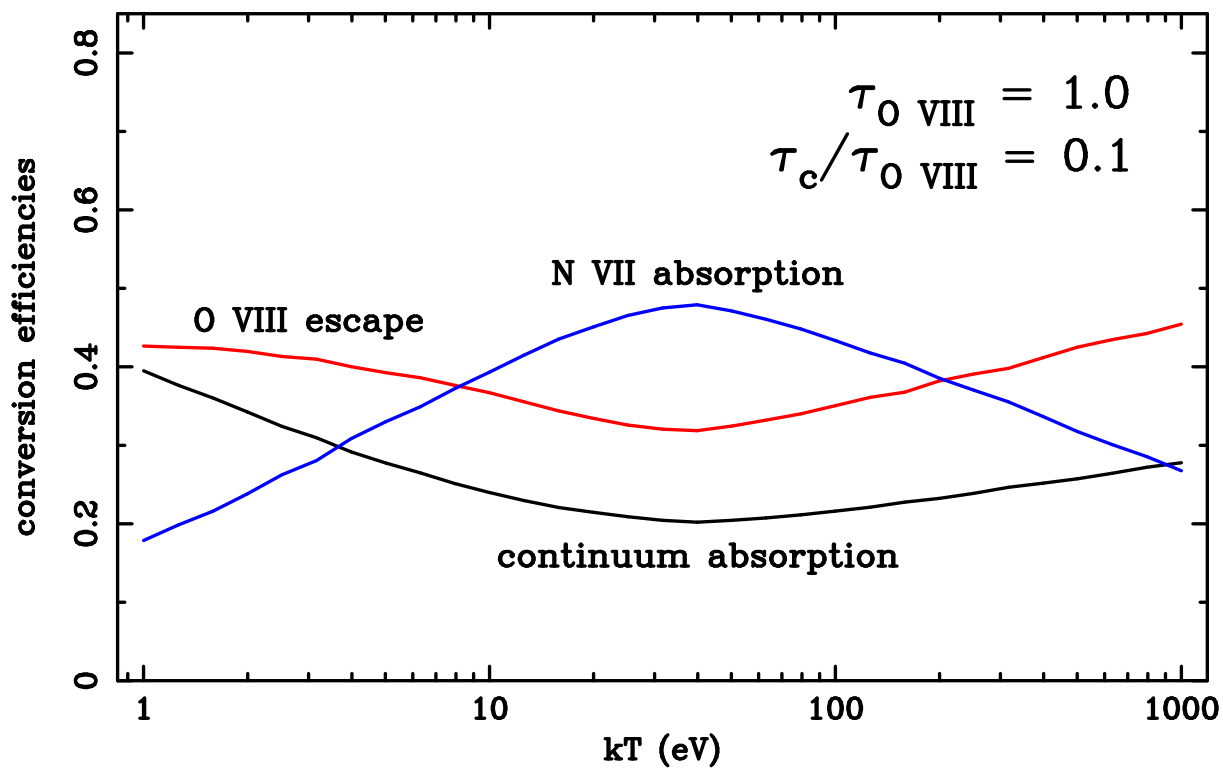}
  \caption{Temperature dependence of the fractional conversion efficiencies.
  The column density is fixed such that $\tau_{\rm{O~VIII}} = 1.0$ and $\tau_c
  = 0.1~\tau_{\rm{O~VIII}}$.  Conversion into \nvii\ peaks at a temperature of
  $kT \sim 40~\rm{eV}$, which corresponds to the temperature where the
  separation between the \oviii\ \lyatwo\ and \nvii\ \lyz\ lines is
  approximately one Doppler width (see text).}  \label{fig:yield_kt}
\end{figure}

\begin{figure}
  \includegraphics[scale=0.7,angle=-90]{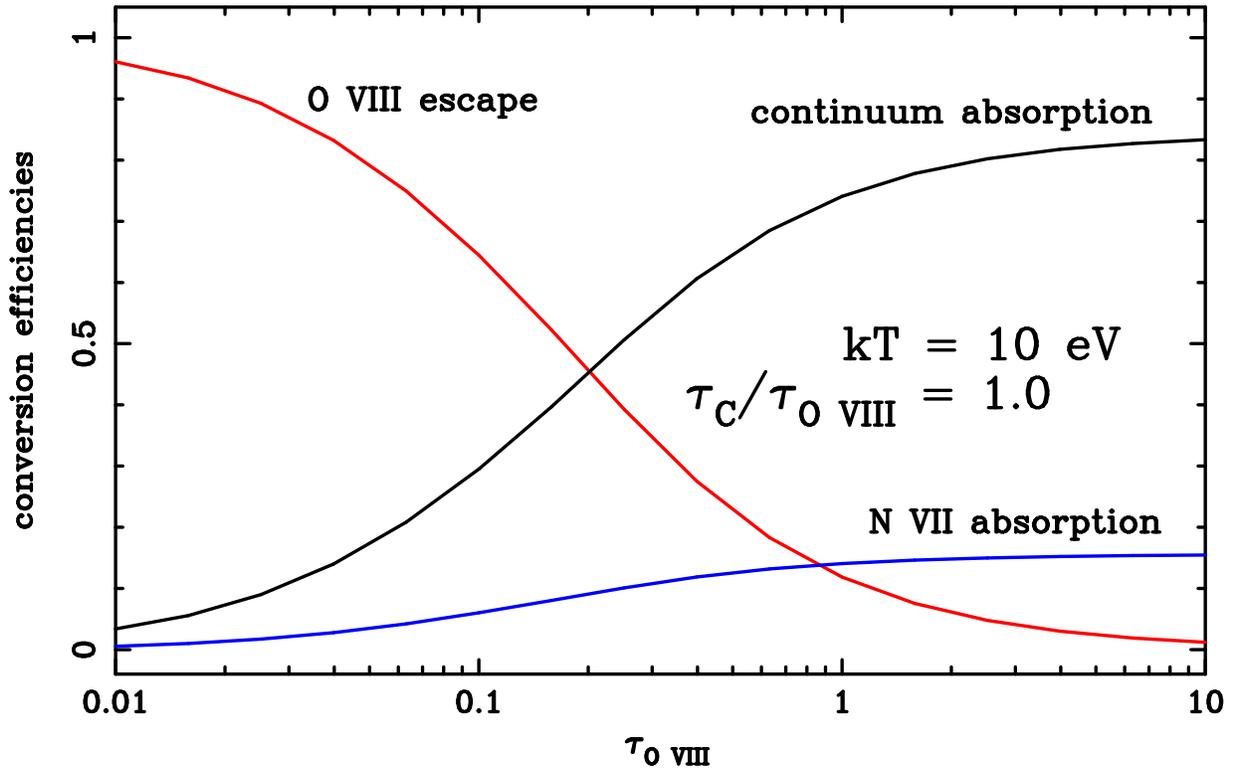}
  \caption{Same as in Figure~\ref{fig:yield10} with the continuum opacity
  increased by a factor of 10 ($\tau_c = \tau_{\rm{O~VIII}}$).  In this case,
  continuum absorption dominates over line absorption by \nvii.}
  \label{fig:yield2}
\end{figure}

  The temperature dependence of the conversion efficiencies can be understood
  from the curves plotted in Figure~\ref{fig:yield_kt}.  At low temperatures,
  the \nvii\ \lyz\ line lies in the wings of the \oviii\ \lya\ lines and so
  the conversion process is relatively inefficient.  As the temperature is
  increased, the amount of overlap between the lines increases rapidly up to a
  temperature of $kT \sim 40~\rm{eV}$, which is where the separation between
  the \nvii\ \lyz\ and \oviii\ \lyaone\ lines is approximately equal to one
  Doppler width.  This temperature is representative of a photoionized plasma
  at this level of ionization.  Above this temperature, the amount of line
  overlap increases more slowly than the temperature dependence of the line
  cross section, which decreases approximately as $\propto T^{-1/2}$, and so
  the efficiency drops again with temperature.  Finally, we show in
  Figure~\ref{fig:yield2} the effect of increasing the overlapping continuum
  opacity on the conversion efficiencies.

  From the derived conversion efficiencies, one can estimate the emergent
  intensities of each of the lines for a given total optical depth.  At $kT =
  10~\rm{eV}$ and a total optical depth of $\tau_{\rm{O~VIII}} = 1$,
  approximately 40\% of the original \oviii\ \lya\ photons escape through the
  surfaces of the slab, 40\% of then are absorbed by \nvii, and the remaining
  20\% are absorbed in the continuum (see Figure~\ref{fig:yield10}).  Assuming
  that the medium is very highly ionized consisting only of H-like carbon,
  nitrogen, and oxygen, and further assuming that the emission coefficient
  ratios of the \lya\ lines are proportional to their abundance ratios, (i.e.,
  \cno\ \lya\ relative line intensities of $1.0:0.13:0.45$), the emergent
  radiation field will have an \oviii\ \lya\ line intensity of $0.40$, a
  \nvii\ \lya\ intensity of $0.26$, and a carbon line intensity of $0.55$.
  Here, we have simply assumed that 33\% of the photons absorbed by \nvii\
  result in the \nvii\ \lya\ line and 50\% of the photons absorbed by \cvi\
  produce the \cvi\ \lya\ line (see \S2).  Therefore, the oxygen to nitrogen
  line ratio decreased from 7.7 to 1.5 (a factor of $\sim 5$) and the oxygen
  to carbon line ratio decreased from 2.2 to 0.73 (a factor of $\sim 3$)
  simply due to finite opacity effects.

  We do not attempt to compute the global emergent spectrum, which requires a
  self-consistent calculation of the \oviii\ \lya\ line source functions
  including a complete set of ions and a realistic ionization balance
  calculation.  To accurately compute the emergent \cno\ line intensities, one
  must properly account for all of the processes that increase the oxygen line
  source function.  The forest of iron L-shell line emission shortward of the
  \oviii\ edge at $\lambda = 14.228$~\AA\ is particularly important and is
  likely to be the dominant contributor to the source function.  A more
  detailed investigation will be presented in a future article.

\section{Conclusions and Discussion}

  We have identified a pair of overlapping resonance lines, which alters the
  intensities of some of the most prominent lines in X-ray spectra of cosmic
  sources under optically thick conditions, analogous to the \ion{He}{2} --
  \ion{O}{3} Bowen fluorescence mechanism.  A non-negligible fraction of the
  \oviii\ \lya\ photons can be converted into \nvii\ emission lines, which can
  be misinterpreted as an anomalous {\cno} abundance pattern.  In general,
  oxygen emission lines will be suppressed, while both nitrogen and carbon
  lines will be enhanced compared to those emitted under optically thin
  conditions.

  As briefly mentioned earlier, detections of non-solar \cno\ abundance ratios
  have been made in a variety of sources using soft X-ray data obtained with
  the Reflection Grating Spectrometer onboard \xmm\ and the \chandra\ Low
  Energy Transmission Grating Spectrometer (see, e.g., \citealt{branduardi01,
  kinkhabwala02, jimenez02, brinkman02, mason03, leutenegger03}).  In all
  cases, the nitrogen lines are observed to be stronger than expected relative
  to the oxygen lines.  The carbon lines, on the other hand, are stronger in
  some sources and weaker in others.  Although it is highly likely that
  non-solar overabundances play a major role in producing the anamolous line
  fluxes, it is also possible that a substantial part of it is due to the
  Bowen mechanism described in this paper.  Detailed modeling and careful
  inspections of other lines in the observed spectrum are required to break
  the degeneracy between overabundance and radiative transfer effects.

  Dynamical effects, such as turbulent motion and velocity gradients, which
  are not included in the present calculations, may reduce or enhance the
  conversion efficiency of \oviii\ \lya\ into \nvii\ depending on the
  geometrical configuration of the ionized medium.  Turbulence, for example,
  can suppress the efficiency if the turbulent length scale is much smaller
  than the mean free path of the \oviii\ line photons.  In the presence of
  velocity gradients, the efficiency can be enhanced in the direction where
  the \nvii\ \lyz\ line is shifted into the \oviii\ \lya\ line core.  These
  effects coupled with density inhomogeneities will complicate the problem
  further and must be studied carefully on a case-by-case basis.

  There are a few other potentially important line overlaps in the X-ray band
  that are worth mentioning.  For example, one of the three brightest
  \ion{Fe}{17} lines at wavelengths of 17.096~\AA, 17.051~\AA, and 16.780~\AA\
  typically referred to as the M2, 3G, 3F lines \citep{brown98}, respectively,
  coincide in wavelength with the high-$n$ series lines of \ovii\ at
  17.10068~\AA\ ($n = 7$), 17.02584~\AA\ ($n = 8$), etc., and 16.786 \AA\ ($n
  = \infty$).  In this case, \ovii\ dominates the line opacity, while
  \ion{Fe}{17} contributes most of the intrinsic source function.  Another
  example is the overlap between the \ion{Fe}{18} lines at 16.071 and
  16.004~\AA\ \citep{brown02} with the \ion{O}{8} \lyb\ lines at 16.00666~\AA\
  and 16.00552~\AA.  These examples are certainly not exhaustive and it is
  highly likely that they must all be accounted for collectively for
  accurately predicting the global X-ray spectrum of optically thick sources.

\acknowledgements The author thanks Roger Blandford, Steven Kahn, and Duane
  Liedahl for valuable discussions, and Mingfeng Gu for his help with the
  Flexible Atomic Code.  This work was supported by \nasa\ through \chandra\
  Postdoctoral Fellowship Award Number {\small PF}1-20016 issued by the
  \chandra\ X-ray Observatory Center, which is operated by the Smithsonian
  Astrophysical Observatory for and behalf of \nasa\ under contract {\small
  NAS}8-39073.

\clearpage

\begin{deluxetable}{ccccc}
  \tablewidth{0pt}
  \tablecaption{Adopted Atomic Parameters\tablenotemark{a} \label{tbl1}}
  \tablehead{
    \colhead{transition}  & \colhead{$\lambda$ (\AA)} & \colhead{$f_{ul}$} &
    \colhead{$A_r$ (s$^{-1}$)} & \colhead{$1 - \epsilon$}
  }
  \startdata
  \ion{O}{8} \lyaone & 18.96711 & $2.77 \times 10^{-1}$ &
     $2.58 \times 10^{12}$ & 1.000 \\
  \ion{O}{8} \lyatwo & 18.97251 & $1.39 \times 10^{-1}$ &
     $2.58 \times 10^{12}$ & 1.000 \\
  \ion{N}{7} \lyz    & 18.97413 & $4.81 \times 10^{-3}$ &
     $2.97 \times 10^{10}$ & 0.797 \\
  \enddata

  \tablenotetext{a}{Wavelengths are adopted from \citet{garcia65} and
  \citet{johnson85}.  All other parameters are calculated with \fac\
  \citep{fac}.}
\end{deluxetable}

\clearpage


\end{document}